\documentclass[aps,prb,twocolumn,superscriptaddress,showpacs]{revtex4-1}
\usepackage{graphicx,color}
\definecolor{red}{rgb}{0,0,0}
\definecolor{green}{rgb}{0,1,0}
\newcommand{\red}{\color{red}}

\begin{document}


\title{Evolution of magnetic fluctuations through the Fe-induced paramagnetic to ferromagnetic transition in Cr$_2$B}

\author{D. Ar\v{c}on}
\email{e-mail: denis.arcon@ijs.si}
\affiliation{Jo\v{z}ef Stefan Institute, Jamova c. 39, 1000 Ljubljana, Slovenia}
\affiliation{Faculty of Mathematics and Physics, University of Ljubljana, Jadranska c. 19, 1000 Ljubljana, Slovenia}

\author{L. M. Schoop}
\affiliation{Department of Chemistry, Princeton University, Princeton, New Jersey 08544, USA}
\affiliation{Max-Planck-Institut f\"{u}r Chemische Physik fester Stoffe, 01187 Dresden, Germany}

\author{R. J. Cava}
\affiliation{Department of Chemistry, Princeton University, Princeton, New Jersey 08544, USA}

\author{C. Felser}
\affiliation{Max-Planck-Institut f\"{u}r Chemische Physik fester Stoffe, 01187 Dresden, Germany}

\date{\today}

\begin{abstract}
 In itinerant ferromagnets, the quenched disorder is predicted to dramatically affect the ferromagnetic to paramagnetic   quantum phase transition {\red driven by external control parameters at zero temperature}. Here we report a study on Fe-doped Cr$_2$B, which, starting from the paramagnetic parent, orders ferromagnetically for Fe-doping concentrations $x$ larger than $x_{\rm c}=2.5$\%. In parent Cr$_2$B, $^{11}$B nuclear magnetic resonance data reveal the presence of both ferromagnetic  and antiferromagnetic fluctuations. The latter are suppressed with Fe-doping, before the ferromagnetic ones finally prevail for $x>x_{\rm c}$. Indications for non-Fermi liquid behavior, usually associated with the proximity of a quantum critical point, were found for all samples, including undoped Cr$_2$B. The sharpness of the ferromagnetic-like transition changes on moving away from $x_{\rm c}$, indicating significant changes in the nature of the magnetic transitions in the vicinity of the quantum critical point. Our data provide constraints for understanding quantum phase transitions in itinerant ferromagnets in the limit of weak quenched disorder.     
\end{abstract}


\maketitle
\section{Introduction}

Itinerant ferromagnets are a class of materials for which the transition from the paramagnetic to ferromagnetic state is considered as a canonical example of a second order phase transition. When an  external control parameter such as pressure suppresses the ferromagnetic state, then these materials approach a quantum critical point\cite{QPT} (QCP) that separates the itinerant ferromagnetic state from its paramagnetic counterpart.\cite{Belitz_PRL2014} 
{\red In such cases, the transition between two nearly degenerate magnetic ground states  is driven by nonthermal fluctuations and the system undergoes a quantum phase transition at zero temperature.}
In an external magnetic field, however, the transition, before being completely suppressed, changes to first order below a tricritical temperature.\cite{Belitz_PRL1999, Belitz_PRL2002} Moreover, while itinerant ferromagnets are usually well described within the framework of standard Fermi liquid theory, in the proximity of a QCP a non-Fermi liquid state can often be inferred\cite{Varma_FL, Vojta_RMP} from resistivity measurements, i.e. the resistivity  displays a power-law temperature dependence $\rho\propto T^{n}$ with an unconventional exponent of $n<2$.\cite{NFL_Nature2001} The existence of a QCP, the quantum critical region at finite temperatures and non-Fermi liquid behavior has been experimentally verified on a number of clean itinerant ferromagnets, such as $3d$-based MnSi,\cite{NFL_Nature2001, MnSi1} ZrZn$_2$ (Ref. \onlinecite{ZrZn2}) or NbFe$_2$ (Ref. \onlinecite{NbFe2}) and $5f$-based UGe$_2$,  URhGe or UCoGe.\cite{UGe1, UGe2, UGe3, UGe4}

In contrast to the established case of QCPs in the clean itinerant ferromagnets mentioned above, the behavior in systems with quenched disorder is less clear. Quenched disorder is predicted to suppress the tricritical temperature until it vanishes at a critical value of disorder and the transition changes back to second order with non-mean-field exponents.\cite{Belitz_PRL2014} However, examples of itinerant ferromagnets where the ferromagnetic transition is suppressed by  quenched disorder at the QCP are remarkably scarce. Therefore, there is a need for new model systems where the nature of the phase transition and possible deviations from the Fermi-liquid state can be systematically investigated close to the QCP in the presence of disorder. 

\begin{figure}[htbp]
\includegraphics[width=1.0\linewidth]{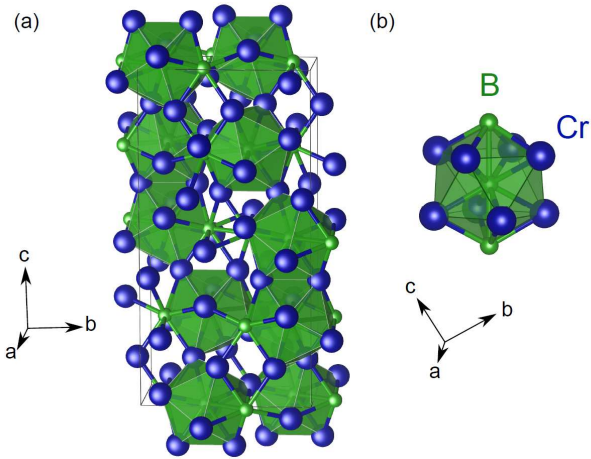}
\caption{\label{fig0} (color online). (a) The orthorhombic crystal structure of Cr$_2$B. (b) The local coordination of each boron site (green polyhedron) has eight nearest Cr atoms  (large blue spheres) and two more B atoms (small green spheres) at apical positions.}
\end{figure}

Recently, a paramagnetic to ferromagnetic phase transition has been reported in lightly Fe-doped Cr$_2$B.\cite{LSchoop} The parent Cr$_2$B is an intermetallic compound that crystallizes in an orthorhombic structure\cite{struct, Struct2} (Fig. \ref{fig0}) with many bands crossing the Fermi energy.\cite{Dirac} Due to the glide planes in the nonsymmorphic space group $Fddd$, the presence of three-dimensional Dirac points in the electronic structure is symmetry dictated and may account for the $n-$type carriers with high mobility measured in Hall effect experiments.\cite{Dirac} According to first principle calculations, the ground state of Cr$_2$B should be antiferromagnetic.\cite{Zhou2009} Experimentally, antiferromagnetic correlations have indeed been inferred from the magnetization data, although no transition to a long-range ordered phase in Cr$_2$B has been found.\cite{LSchoop} Upon Fe-doping, detailed Arrott analysis of the temperature and the field dependences of magnetization data revealed that a ferromagnetic phase emerges at a critical Fe-concentration near $x=0.02$.\cite{LSchoop} Moreover, the observed logarithmic contribution to the heat capacity was taken as a hallmark of quantum criticality, and the power-law exponent $n\sim 1$ in the temperature dependence of resistivity was taken to indicate a non-Fermi liquid state in doped samples. However, whether Fe-doped Cr$_2$B does indeed represent a new family of intermetallics where quenched disorder drives a ferromagnetic quantum phase transition calls for additional independent experimental confirmation.   

Magnetic resonance techniques, such as nuclear magnetic resonance (NMR) and electron spin resonance (ESR), have proven to be extremely powerful probes of local spin susceptibilities and spin fluctuations close to a QCP.\cite{CFNMR, CFNMR1,CFNMR2, CFNMR3, CFNMR4, ESR, ESR1} Here we report  detailed $^{11}$B NMR and ESR measurements on polycrystalline Fe-doped Cr$_2$B at different doping levels across the paramagnetic to ferromagnetic transition. Surprisingly, we find a strong temperature dependent shift of the $^{11}$B NMR spectra that generally follows a Curie-Weiss-like dependence, revealing a crossover from predominantly antiferromagnetic correlations to predominantly ferromagnetic correlations at a critical Fe-doping level of $x\sim 2.5$\%. On cooling, we find a crossover to a low-temperature state where the $^{11}$B NMR shift displays a non-Fermi-liquid $T^{-1/2}$-dependence, thus complying with what is expected in the vicinity of a QCP. Simultaneously, the appearance of a strong electron spin resonance signal and a substantial broadening of the $^{11}$B NMR spectra provide evidence for ferromagnetic-like ordering at low-temperatures that is completely absent at lower Fe-doping levels. The data presented here reveal highly unusual behavior for Fe-doped Cr$_2$B, which cannot be simply rationalized within a standard Fermi-liquid model. Moreover, the systematic evolution of the sharpness of the ferromagnetic-like transition with Fe-doping level implies significant changes in the nature of the phase transition as the quenched disorder varies across the critical Fe-doping value for the quantum phase transition (QPT).

\section{Experimental methods}

For this study, the polycrystalline Fe-doped Cr$_2$B samples used in Ref. \onlinecite{LSchoop} to characterize the magnetic and transport properties of the Cr$_{2-x}$Fe$_{x}$B system were employed. The Fe-doping range is between 0 and 5\%. According to powder x-ray diffraction,
only very small fractions of non-magnetic Cr metal were identified in the diffraction profiles in addition to the doped Cr$_2$B phase. The samples were also characterized in detail by high resolution transmission electron microscopy and high angle annular dark-field scattering transmission microscopy to exclude the possibility of Fe clustering. The magnetic susceptibility data excluded the presence of magnetic impurities.\cite{LSchoop} 
 
For the temperature-dependent ESR experiments, the samples were sealed under helium in a standard 4 mm-diameter silica tube (Wilmad Lab Glass) whereas for the $^{11}$B NMR experiments, samples with a  mass of around 90~mg were directly inserted into the NMR coil. A conventional continuous wave (cw) electron paramagnetic resonance spectrometer operating at a Larmor frequency $^{\rm ESR}\nu_{\rm L}=9.6$~GHz was employed to detect the electron spin resonance. The spectrometer is equipped with a standard Varian E-101 microwave bridge, a Varian rectangular TE102 resonance cavity, and an Oxford Cryogenics continuous-flow helium cryostat. The temperature stability was better than $\pm 0.1$~K over the entire temperature range of measurements ($4-300$~K).

The $^{11}$B (nuclear spin $ I = 3/2$) NMR spectra and the spin-lattice relaxation rates were measured between 5 and 300 K in a magnetic field of $4.7$~T. The $^{11}$B NMR shifts are determined relative to the Larmor frequency $^{11}\nu_{\rm L} = 64.167$~MHz, defined by a BF$_3$Et$_2$O standard. For $^{11}$B NMR line shape measurements, a solid-echo pulse sequence, $\pi/2 - \tau - \pi/2 - \tau - {\rm echo}$, was employed, with a pulse length $t_w(\pi/2) = 1.9$~$\mu$s and an interpulse delay $\tau = 40$~$\mu$s. The complete polycrystalline
NMR spectrum was obtained by summing the real part of spectra measured step-by-step at resonance frequencies separated by $\Delta\nu = 50$~kHz. Since it was impossible to completely  invert the $^{11}$B nuclear magnetization, we used the saturation-recovery pulse sequence for the spin-lattice relaxation rate measurements. Typically, the saturation train consisted of a sequence of 20 $\pi/2$ pulses separated by 50 $\mu$s.

Because $^{11}$B is a quadrupole nucleus, we model the $^{11}$B NMR spectra with the general spin Hamiltonian ${\cal H}={\cal H}_{\rm Z}+{\cal H}_{\rm Q}+{\cal H}_{\rm S}$ comprising the nuclear Zeeman (${\cal H}_{\rm Z}$), nuclear quadrupole (${\cal H}_{\rm Q}$) and $^{11}$B shift (${\cal H}_{\rm S}$) terms, respectively. 
The later gives rise to the $^{11}$B NMR lineshift of the central $- 1/2 \leftrightarrow 1/2$ transition, $K$, and  includes the temperature independent chemical shift, $\sigma$, and the Knight shift, $K_{\rm s}$, which  originates from the hyperfine coupling to itinerant electrons. 
The most important term for our study  is $K_{\rm s}$, as it is directly proportional to the local electronic susceptibility $\chi$, e.g., its isotropic part is given by $K_{\rm iso}={a_{\rm s}\over N_{\rm A}\mu_0}\chi$, where $a_{\rm s}$ is a hyperfine coupling constant, $N_{\rm A}$ is Avogadro's number and $\mu_0$ is the magnetic permeability of vacuum.  When analyzing the $^{11}$B NMR spectra of Cr$_2$B we include the  $^{11}$B anisotropic shift  and the quadrupole effects up to the second order.  Fitting of the spectra thus yields, in addition to $K$,  the quadrupole splitting frequency $\nu_Q = {3eV_{zz}Q \over h2I(2I-1)}$  and the asymmetry parameter $\eta = (V_{xx}-V_{yy})/V_{zz}$, both defined by the components of the electric field gradient (EFG) $V_{ij}$ at the $^{11}$B site. Here $Q$ and $h$ are the $^{11}$B quadrupole moment and the Planck constant, respectively. The powder NMR spectrum is computed as a histogram of resonance frequencies obtained by summing over uniformly distributed polar and azimuthal angles of the
magnetic field orientation with respect to the quadrupole tensor principal axes.\cite{Tone_La2C3, NOFA} We  included the homogeneous broadening of the line through the convolution of the computed spectra with a Lorentzian line with a full-width-half-maximum $\delta_{1/2}$.  The effect of  quadrupole splitting frequency distribution on $^{11}$B NMR spectra  was also tested by using normally distributed values of quadrupole frequencies with the center at $\nu_Q$ and width of $\Delta\nu_Q$.

\begin{figure}[t]
\includegraphics[width=1.0\linewidth]{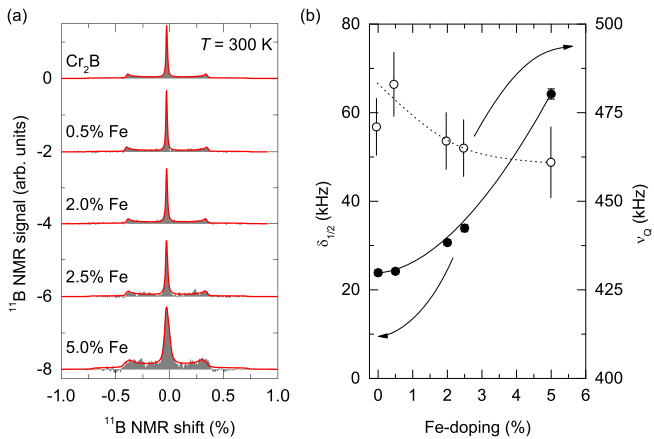}
\caption{\label{fig1} (color online). (a) Comparison of the room-temperature $^{11}$B NMR spectra (gray shaded area) measured on polycrystalline Fe-doped Cr$_2$B samples at various Fe-doping levels. Solid red lines are the powder lineshape fits to a model that includes anisotropic shift and quadrupole effects up to the second order. A single $^{11}$B NMR component was sufficient in all cases. (b) Fe-doping dependence of the $^{11}$B NMR  linewidth broadening parameter, $\delta_{1/2}$, (solid circles, left scale) and of the quadrupole frequency  $\nu_{Q}$, (open circles, right scale). The solid and dashed black lines are guides to the eye for the Fe-doping dependence of $\delta_{1/2}$ and $\nu_{Q}$, respectively.}
\end{figure}
%


\section{Results and Discussion}

\subsection{Homogeneity of the Cr$_2$B samples after Fe doping}

The room temperature $^{11}$B NMR spectrum of Cr$_2$B sample [Fig. \ref{fig1}(a)] displays a characteristic quadrupole powder lineshape with a clearly pronounced satellite transition ($\pm 3/2 \leftrightarrow \pm 1/2$) singularities flanking the central peak that corresponds to a $- 1/2 \leftrightarrow 1/2$ transition. The spectrum is considerably shifted by $K=-228(3)$~ppm relative to the $^{11}$B NMR reference frequency. The central peak has a nearly Lorentzian lineshape thus implying that the broadening due to the anisotropic shift interactions and  the second-order quadrupole corrections is negligible.  A  lineshape fit of the spectrum yields a quadrupole splitting frequency of $\nu_Q = 471(8)$~kHz  and an asymmetry parameter of $\eta = 0.02(1)$. Surprisingly, we find $\eta\approx 0$, although the $^{11}$B site symmetry -- B is at the low-symmetry  $16g$ (0.125, 0.125,  0.4993) site [Fig. \ref{fig0}(b)]\cite{struct, Struct2} -- does not require such a restriction. The homogeneous broadening is $\delta_{1/2} =24(1)$~kHz.  

Light doping of Cr$_2$B with 0.5\% Fe induces almost no change to the $^{11}$B NMR lineshape [Fig. \ref{fig1}(a)] and the unconstrained fit returns, compared to parent undoped Cr$_2$B, nearly the same $\nu_Q=483(9)$~kHz, $\eta=0.02(1)$ and $\delta_{1/2} =24(1)$~kHz. As the level of Fe-doping increases, however, the $^{11}$B NMR spectra drastically broaden, and for  5\% Fe doping the central peak also becomes slightly anisotropic. On extracting the parameters from the $^{11}$B NMR lineshape fits, we  first notice that $\nu_Q$  marginally reduces with  increasing Fe concentration [Fig. \ref{fig1}(b)], i.e. to 467(8)~kHz, 465(8)~kHz and 461(10)~kHz in the 2\%, 2.5\% and 5\% doped samples, respectively. This insensitivity of $\nu_Q$ to the Fe-doping implies that the lattice around the B-sites contributing to the measured spectra is not markedly perturbed for Fe-doping levels up to 5\%. (We note that it is possible that the B atoms sitting next to Fe-dopant atoms experience significantly different EFG and hyperfine fields that shift and broaden the spectra beyond the sensitivity of the present NMR experiments.)  Nevertheless, the local hyperfine fields at ``weakly perturbed'' $^{11}$B sites do change, judging from the   Fe-doping variation in the linewidth parameter $\delta_{1/2}$, which  increases to 31(1)~kHz, 34(1)~kHz and 64(2)~kHz for 2\%, 2.5\% and 5\% Fe doping, respectively [Fig. \ref{fig1}(b)]. Simultaneously, satellite transition singularities become significantly less pronounced, which explains the larger $\eta = 0.08(1)$ for the largest Fe-doping level. (We note that for the 5\% doped sample a $^{11}$B NMR lineshape fit that includes a small distribution of $\nu_Q$, i.e. $\Delta\nu_Q /\nu_Q =6$\%, describes the experimental spectrum equally well.) Therefore we find that Fe-doping indeed introduces some local-site disorder, which is rather sensitively picked up by the NMR parameters. However, because all $^{11}$B NMR spectra can still be simulated with a single well-defined value of $\nu_Q$, we conclude that the Fe-doping must be rather homogeneous for all samples, consistent with previous chemical and structural characterization.\cite{LSchoop}

\subsection{The paramagnetic state of parent Cr$_2$B}

\begin{figure}[t]
\includegraphics[width=1.0\linewidth]{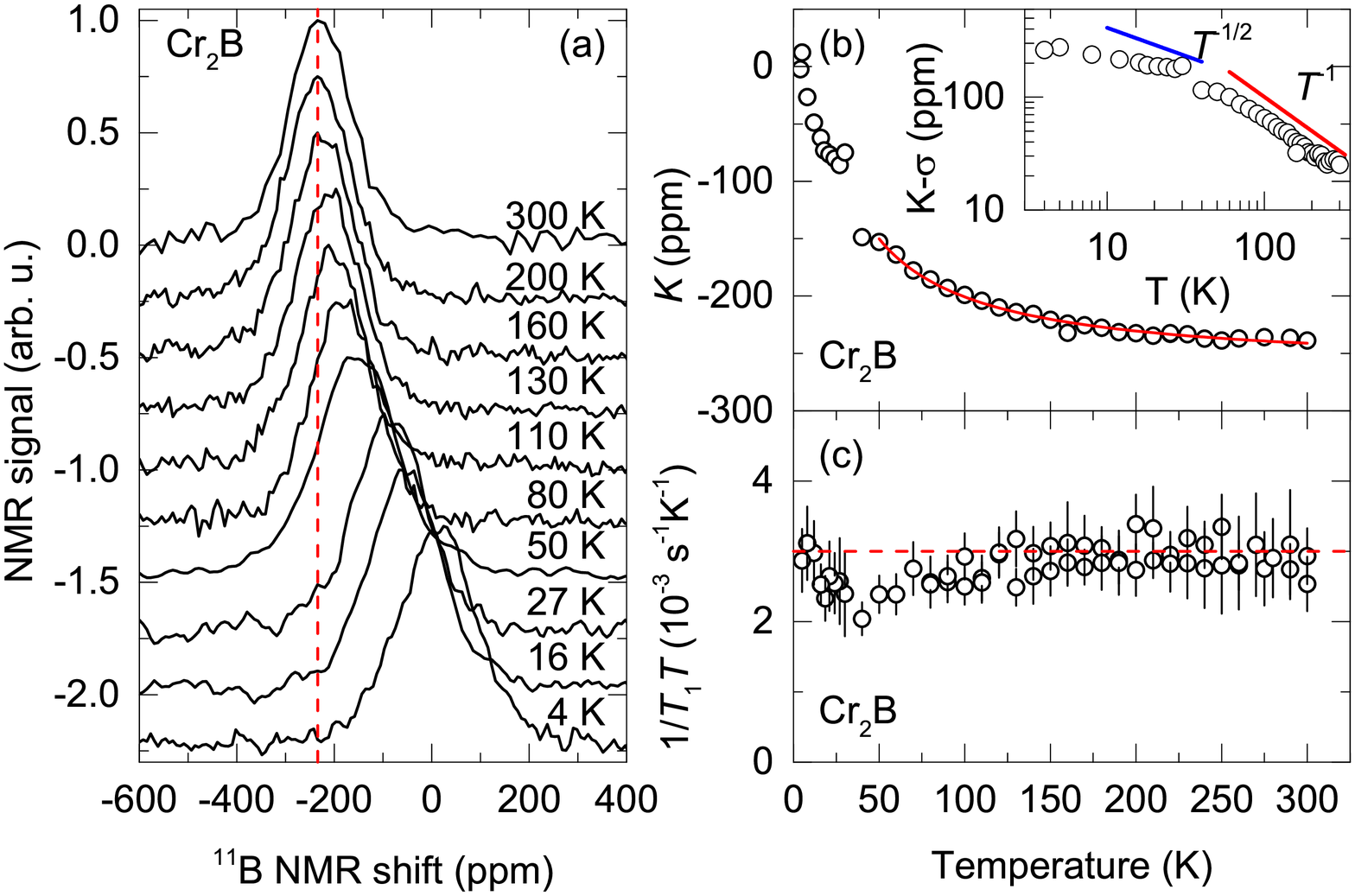}
\caption{\label{fig2} (color online). Temperature dependences of (a) the $^{11}$B central transition ($-1/2\leftrightarrow 1/2$) peak, (b) the shift of the central transition peak, $K$, and the spin-lattice relaxation rate $1/T_1$ divided by $T$ measured in parent undoped Cr$_2$B. The vertical and horizontal dashed lines in (a) and (c) mark room temperature values. The solid line in (b) is a fit with a Curie-Weiss dependence to a negative Curie-Weiss temperature of $T_{cw}=-12(3)$~K. The inset to (b) shows the temperature dependence of the Knight shift, $K_{\rm s}$, on a log-log scale, obtained after subtracting the chemical shift $\sigma$ from $K$. Solid lines indicate the high-temperature slope of $K_{\rm s}\propto T^{-1}$ and the low-temperature slope of $K_{\rm s}\propto T^{-1/2}$.  }
\end{figure}

Fig. \ref{fig2}a shows the temperature dependence of the central transition peak in parent Cr$_2$B. This peak retains its Lorentzian lineshape at all temperatures and shows almost no broadening between room temperature and $T=4$~K. The complete absence of broadening of the powder $^{11}$B NMR spectra is a firm evidence that no static magnetic order is established.  However, the spectra monotonically shift towards higher frequencies with decreasing temperature. Between room temperature and 50 K the $^{11}$B NMR shift $K$ shows [Fig.  \ref{fig2}(b)] a  Curie-Weiss-like dependence 
\begin{equation}
K(T)=\sigma+B/(T-T_{cw})\, .
\label{CW}
\end{equation}  
Here we identify the constant $\sigma=-264(3)$~ppm as a temperature independent  chemical shift. On the other hand, the second temperature-dependent Knight-shift contribution originates from the hyperfine interactions with the unpaired electronic moments. The extracted negative Curie-Weiss temperature $T_{cw}=-12(3)$~K implies that the involved itinerant electronic states are antiferromagnetically correlated and thus corroborates the magnetization data\cite{LSchoop} {\red and the first principle calculations.\cite{Zhou2009}}

At $T\approx 35$~K we notice a small discontinuity in $K$, but for lower temperatures  $K$ still continues to increase with decreasing temperature. Plotting the temperature dependence of the Knight shift $K_{\rm s}(T)=K(T)-\sigma$ on a log-log plot [inset to Fig. \ref{fig2}(b)], we notice a gradual change of slope at around 35~K. Whereas at higher temperatures the slope of $K_{\rm s}$  indeed fits to a $T^{-1}$ dependence, as established above, we find that for lower temperatures $K_{\rm s}$ develops approximately a $T^{-1/2}$ dependence before leveling off at the lowest temperatures. A very similar sequence of power-laws in the temperature dependence of the Knight shift has been reported for YbRh$_2$Si$_2$,\cite{CFNMR} and attributed to non-Fermi-liquid behavior in the vicinity of a QCP. 

Next, the spin dynamics in the paramagnetic state of parent Cr$_2$B were probed through the $^{11}$B spin-lattice relaxation rate $1/T_1$. In striking contrast to the strongly temperature dependent $K$, the spin-lattice relaxation rate divided by temperature, $1/T_1T$, shows a much weaker temperature dependence [Fig.  \ref{fig2}(c)]. Between room temperature and 35~K, $1/T_1T$ gradually decreases by $\sim 30$\%, but then on further cooling it starts to increase again. 

For correlated metals where electron-electron exchange enhancement effects are important, the Korringa relation is \cite{Walstedt}
\begin{equation}
{{}^{11}T_1TK_{\rm s}^2}= {\hbar \over 4\pi k_{\rm B}} {\gamma_{\rm e}^2\over \gamma_{11}^2}\beta\, .
\label{Korr}
\end{equation}
Here $\gamma_{\rm e}$ and $\gamma_{11}$ are the electronic and $^{11}$B  gyromagnetic ratios, respectively. The  Korringa factor $\beta$ is introduced to account for the electron-electron exchange.\cite{Walstedt, peter, Stenger} Inserting the room-temperature value of $1/T_1T =2.9\cdot 10^{-3}$~${\rm s}^{-1}{\rm K}^{-1}$ and $K_{\rm s}=25$~ppm into Eq. (\ref{Korr}) we calculate $\beta = 3.3$ thus complying with the enhancement of dynamic spin susceptibility. In general, $\beta > 1$ implies ferromagnetic fluctuations,\cite{peter, Stenger} seemingly contradicting the analysis of the Knight-shift data, which disclosed the presence of antiferromagnetic  correlations. 
{\red We additionally note that, although the precise value of  the Korringa factor $\beta$ depends on the choice of $\sigma$, we still find that the increase of $\beta$ with decreasing temperature further corroborates the presence of ferromagnetic correlations.}
In order to explain this apparent contradiction, we suggest that  both ferromagnetic and antiferromagnetic spin fluctuations are present, but the later are filtered out  at the $^{11}$B site.
{\red Namely, the local coordination of  $^{11}$B site has eight nearest neighboring Cr atoms and two more  $^{11}$B sites at the apical positions [Fig. \ref{fig0}(b)].  We next point out that near the Fermi level, the calculated density of states is dominated by Cr $d$ bands.\cite{Zhou2009}  
Therefore, the appropriate electron-nuclear Hamiltonian consists of a sum of transferred hyperfine coupling of $^{11}$B at site $k$ to the eight nearest neighbor Cr electron spins at $k’$. As a result, the contribution of antiferromagnetic spin fluctuations to the spin-lattice relaxation may be suppressed.\cite{Moriya, Walstedt}} The important conclusion from this  part of our study is thus that although the multi-band Cr$_2$B parent compound displays both antiferromagnetic and ferromagnetic correlations, they are not sufficiently strong to establish a long-range magnetically ordered state. The observed anomalies at $\sim 35$~K in both $K(T)$ and $1/T_1T$ data imply the presence of subtle electronic changes that may preclude magnetic ordering at low temperatures, but additional more detailed experiments in this temperature range are required to unveil the nature of the undoped Cr$_2$B system.

\subsection{The ferromagnetic state in Fe-doped Cr$_2$B}

\begin{figure}[t]
\includegraphics[width=0.9\linewidth]{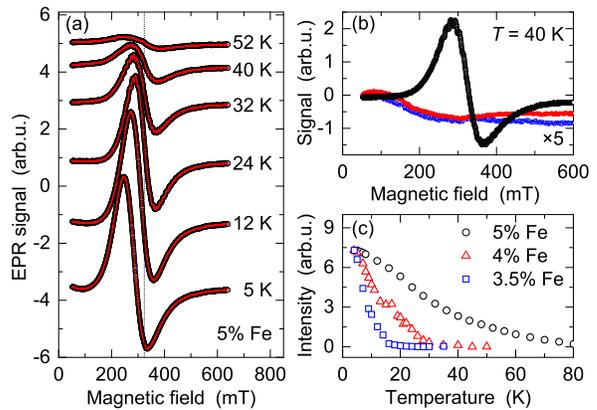}
\caption{\label{fig4} (color online). (a) Temperature dependence of ferromagnetic resonance in 5\% Fe doped Cr$_2$B (black circles) measured at X-band ($^{\rm ESR}\nu_{\rm L}=9.6$~GHz) frequencies. Solid red lines are fits of the spectra to the Dyson lineshape. The dotted vertical line at 330~mT marks the resonance field of the $g=2$ electron paramagnetic resonance signal. (b) Comparison of ferromagnetic resonance spectra measured at 40~K for 5\% (black circles), 4\% (red triangles) and 3.5\% (blue squares) Fe-doped Cr$_2$B, respectively. (c) Temperature dependence of the intensities of the ferromagnetic resonance spectra for  5\% (black circles), 4\% (red triangles) and 3.5\% (blue squares) Fe-doped Cr$_2$B, respectively.}
\end{figure}

Cr$_2$B becomes ferromagnetic when 2\% or more of the Cr atoms are replaced by Fe.\cite{LSchoop} In order to investigate the ferromagnetic state we first employ the electron spin resonance technique. For 5\% Fe-doped Cr$_2$B in the high temperature metallic paramagnetic phase, no conducting electron spin resonance (CESR) signal could be detected. However, as the temperature is decreased below $\sim 70$~K, an asymmetric Dyson-like resonance\cite{Dyson} appears at a $g\approx 2$ resonance field [Fig. \ref{fig4}(a)]. The Dyson lineshape of the resonance is a hallmark of a metallic state, thus ruling out previously undetected insulating iron-oxide impurities as a possible source of the observed ESR signal. With decreasing temperature, the signal gradually grows in intensity and shifts to lower resonance fields, signaling the onset and growth of internal magnetic fields in the material. The observed behavior is consistent with what is seen for ferromagnetic resonance in the metallic state, and thus agrees with the proposed ferromagnetic ground state for 5\% Fe-doped Cr$_2$B. 

As the Fe-doping level decreases, the appearance of the ferromagnetic-like resonance is systematically shifted to lower temperatures. For instance, whereas at 40~K in 5\% Fe-doped sample the resonance signal is very strong, no such line was detected at the same temperature for the 4\% and 3.5\% Fe-doping levels [Fig. \ref{fig4}(b)]. It is, however, detected at lower temperatures for those compositions. In order to quantitatively follow the ferromagnetic-like resonance, we next fit the spectra to a Dyson lineshape\cite{Dyson}  [Fig. \ref{fig4}(a)]. The extracted temperature dependences of the intensity of the resonance peak, which is proportional to the sample's magnetization, are summarized for different Fe-doping levels in Fig. \ref{fig4}(c). The magnetic ordering onset temperature $T_{\rm c}$ systematically decreases from $\sim 70$~K at 5\% doping to $\sim 30$~K and then to $\sim 18$~K for 4\% and 3.5\% Fe-doped samples, respectively. In addition, whereas the transition in the 5\% doped sample is smeared over a large temperature interval, it is much sharper at lower doping-levels, which are closer to the QCP. This experimental observation may imply significant changes in the nature of magnetic transition in the vicinity of the QCP.

\begin{figure}[t]
\includegraphics[width=1.0\linewidth]{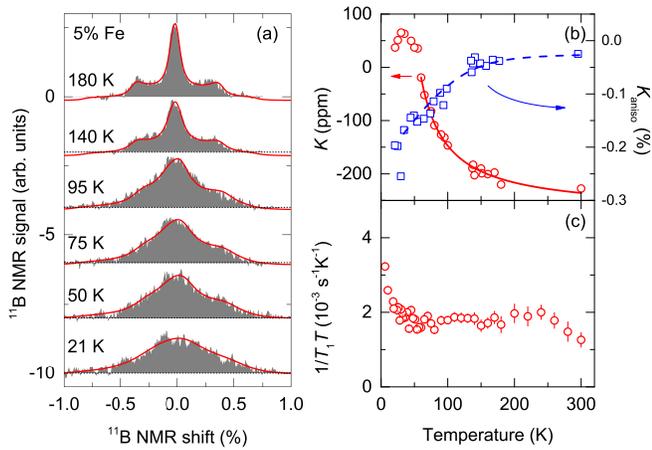}
\caption{\label{fig3} (color online). (a) The temperature evolution of the $^{11}$B NMR spectra for 5\% Fe-doped Cr$_2$B. The solid red lines are fits to a model with quadrupole and anisotropic Knight-shift interactions. In the model, we assumed temperature independent $\nu_{\rm Q}=461$~kHz and $\eta =0.08$. (b) The temperature dependences of the isotropic (red circles, left scale) and anisotropic (blue squares, right scale) parts of the Knight shift. A fit of the $^{11}$B NMR shift, $K(T)$, to Eq. \ref{CW} (solid red line) yields the chemical shift $\sigma=-270(8)$~ppm, $B=9.7(9)\cdot 10^3$~ppm K and the ferromagnetic Curie-Weiss temperature $T_{\rm CW}=21(4)$~K. The dashed blue line is a guide to the eye.  (c) Temperature dependence of the $^{11}$B NMR spin-lattice relaxation rate divided by temperature, $1/T_1T$.}
\end{figure}

Further insight into the development of local magnetic fields is provided by the $^{11}$B NMR spectra measured for 5\% Fe-doped Cr$_2$B [Fig. \ref{fig3}(a)]. 
{\red Whereas the spectra retain a characteristic quadrupole $I=3/2$ powder lineshape with small anisotropic Knight shift interaction [e.g., similar as at 300 K (Fig. 2)] on cooling down to $\sim 140$~K, broadening due to the anisotropic Knight shift interaction gradually begins to dominate the spectra at lower temperatures.}
Below $\sim 50$~K the spectra already display a lineshape that is reminiscent of  anisotropic Knight shift interactions. 
{\red Alternatively, the lineshape broadening originating from the distribution of Knight shifts is less probable because it is not accompanied also by the large  quadrupole frequency distribution.}
Therefore, for the fits to the data, we assumed temperature independent $\nu_{\rm Q}=461$~kHz and $\eta =0.08$, both extracted from the room temperature spectra, and that the isotropic $K$ and anisotropic $K_{\rm aniso}$ parts of the Knight-shift tensor are temperature dependent parameters. The temperature dependences of $K$ and $K_{\rm aniso}$ thus obtained are summarized in Fig. \ref{fig3}(b). Both parameters exhibit a very strong temperature dependence that can again be described as a Curie-Weiss-like dependence between room temperature and 70~K. Fitting the isotropic shift $K(T)$ to Eq. (\ref{CW}) yields $\sigma=-270(8)$~ppm, which is nearly identical to the corresponding chemical shift of the parent Cr$_2$B. On the other hand, we now find a positive Curie-Weiss temperature, $T_{\rm CW}=21(4)$~K, which  is thus fully consistent with the dominant ferromagnetic correlations in heavily Fe-doped Cr$_2$B. The temperature dependence of $K_{\rm aniso}$ also supports this finding. 

Large broadening of the $^{11}$B NMR spectra at low temperatures, reflected in the enhanced $K_{\rm aniso}$, clearly evidences the development of static local magnetic fields and thus of spin-freezing. The ferromagnetic $T_{\rm CW}$ suggests that the magnetic moments induced by Fe-doping freeze into a spin state where ferromagnetic correlations prevail. However, what may be surprising is that the shift of $K(T)$ remains relatively small, i.e. on the order of $\sim 300$ ppm, which is thus only a fraction of $K_{\rm aniso}$. We thus conclude that the contact hyperfine (or the isotropic part of the transferred hyperfine) interaction with itinerant electrons is nearly the same in Fe-doped Cr$_2$B compared to parent undoped Cr$_2$B. The reason for this is currently unknown, but one of the possibilities is that when Fe is introduced into the lattice it creates localized states that interact with $^{11}$B mostly via long-range, e.g. dipolar, interactions. This may also explain the temperature dependence of the spin-lattice relaxation rate [Fig. \ref{fig3}(c)]. Compared to parent Cr$_2$B, the slightly shorter $T_1$ at room temperature yields $1/T_1T= 1.3\cdot 10^{-3}$~${\rm s}^{-1}{\rm K}^{-1}$, and therefore a ferromagnetically enhanced $\beta\approx 8$. On cooling below $\sim 70$~K, $1/T_1T$ indeed starts to increase as expected when close to the magnetic ordering. However, the absence of a divergence in $1/T_1T$, normally found at the magnetic ordering temperature, and the broadening of the ferromagnetic-like transition observed in the ESR data (Fig. \ref{fig4}) may be signatures of a distribution of ferromagnetic freezing temperatures, and thus of a smeared transition to a ferromagnetic-like state with a high degree of disorder.

\subsection{Behavior close to the critical Fe doping composition}

\begin{figure}[t]
\includegraphics[width=1.0\linewidth]{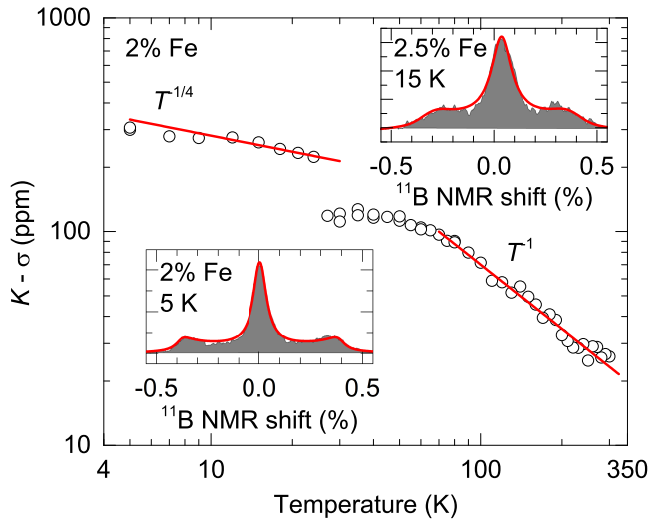}
\caption{\label{fig5} (color online). The temperature dependence of the $^{11}$B NMR Knight shift, $K_{\rm s}=K-\sigma$, for 2\% Fe-doped Cr$_2$B (open circles). The Knight shifts are extracted directly from the shifts of the $^{11}$B NMR spectra, $K$, obtained by subtraction of the chemical shift $\sigma= -253$~ppm. The solid red lines indicate that $K_{\rm s}\propto T^{-1}$ at high temperatures, and then gradually changes to $K_{\rm s}\propto T^{-1/4}$ at low temperatures. Insets: Low-temperature $^{11}$B NMR spectra (gray shaded area) measured for 2\% Fe-doped Cr$_2$B ($T=5$~K) and 2.5\% Fe-doped Cr$_2$B ($T=15$~K). The solid red lines are lineshape fits with $K=54$~ppm, $\nu_{\rm Q}=496$~kHz and $\delta_{1/2}=155$~kHz (Cr$_2$B sample with 2\% Fe doping) and $K=167$~ppm, $\nu_{\rm Q}=450$~kHz and $\delta_{1/2}=243$~kHz (Cr$_2$B sample with 2.5\% Fe doping). }
\end{figure}

Finally, we focus on Fe-doping levels lower than 2.5\%, i.e. the samples {\red with suppressed ferromagnetic ordering temperature being thus} close to the QCP.\cite{LSchoop} Inspecting the low-temperature $^{11}$B NMR  spectra of these samples we find that they retain a characteristic powder quadrupole lineshape at all temperatures (insets to Fig. \ref{fig5}). This proves the absence of the strong internal fields that would broaden the $^{11}$B NMR spectra, as was the case for ferromagnetic 5\% Fe-doped Cr$_2$B  at higher temperatures. The absence of the ferromagnetic-like resonance signal in these samples down to 4~K provides additional evidence for the absence of any magnetic order. On the other hand, the temperature dependence of $K(T)$ is a strong indication of a non-Fermi liquid state. Namely, between room temperature and $\sim 70$~K the shift follows a Curie-Weiss-like dependence (Fig. \ref{fig5}), which seems to be a general characteristic of all the Cr$_2$B samples studied, both doped and undoped. 

Fitting $K(T)$ in the temperature interval between 300 and 70~K reveals a significant Fe-doping dependence of the NMR-determined Curie-Weiss temperature. While we find a small but positive $T_{\rm CW}=8(3)$~K for 2.5\% doped samples and above, it is negative (antiferromagnetic), $T_{\rm CW}=-20(3)$~K, for the 2\% doped material and remains negative to lower Fe-dopeing levels. Similar to the case for parent Cr$_2$B, at lower temperatures $K_{\rm s}(T)$ increases with a smaller power-law exponent, i.e. roughly as $T^{-1/4}$, before leveling off at lowest temperatures. 

Compared to parent Cr$_2$B, the anomaly in the temperature dependence of $K(T)$ shifts from 35~K to 24~K in the 2\% Fe-doped sample (Fig. \ref{fig5}). Surprisingly, this anomaly is not clearly observed in the spin-lattice relaxation rates, which are temperature independent, i.e. $1/T_1T=2.9\cdot 10^{-3}$~${\rm s}^{-1}{\rm K}^{-1}$, between 300 and 5~K. In contrast to the 5\% Fe-doped sample, there is no enhancement in $1/T_1T$ that would suggest the development of a ferromagnetic-like state at low temperatures. The $^{11}$B NMR shift and the spin-lattice relaxation rate data thus unambiguously show that Fe-doped Cr$_2$B materials at concentrations lower than the critical 2.5\% value lack magnetic order, while systematically showing electronic characteristics that deviate from conventional Fermi-liquid behavior.

\section{Discussion and conclusions}

The insensitivity at room-temperature of the quadrapole frequencies $\nu_{\rm Q}$ [Fig. \ref{fig1}(b)] and $^{11}$B NMR shifts [Fig. \ref{fig1}(a)] clearly demonstrates that Fe doping of Cr$_2$B at levels up to 5\% does not significantly perturb the local structural and electronic environment at the B-sites. 
{\red Yet, the emergence of the ferromagnetic-like resonance (Fig. \ref{fig4}) and the large broadening of the $^{11}$B NMR spectra (Fig. \ref{fig3}) observed in 5\% Fe-doped Cr$_2$B are consistent with the presence of magnetic ordering at low temperatures  when the Fe concentration exceeds the critical value of $x_{\rm c}\approx 2.5$\%.} 
{\red Therefore, the quantum paramagnetic to ferromagnetic transition at zero temperature is  indeed triggered by Fe doping at $x_{\rm c}$.}
 So, what really changes after Fe-doping that drives such transition? Our $^{11}$B NMR and ESR data highlight two primary factors that constrain the discussion of the paramagnetic to ferromagnetic transition in this material: the non-Fermi-liquid behavior and the simultaneous presence of antiferromagnetic and ferromagnetic correlations. 

\begin{figure}[t]
\includegraphics[width=1.0\linewidth]{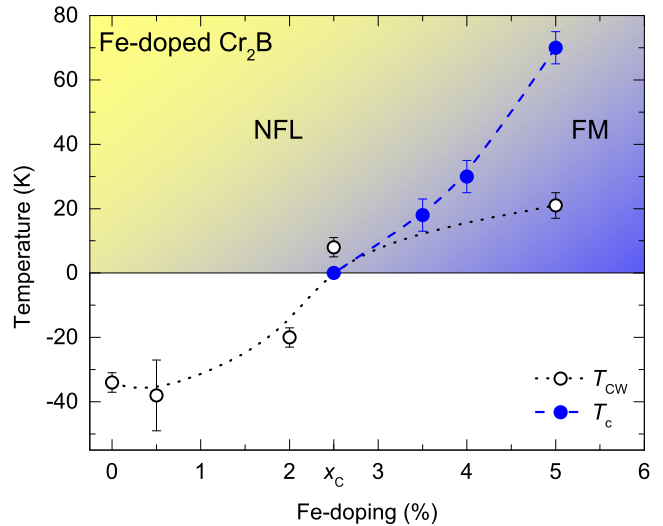}
\caption{\label{fig6} (color online). Dependence of the NMR-determined Curie-Weiss temperature $T_{\rm CW}$ (open black circles) and magnetic ordering onset temperature $T_{\rm c}$ (solid blue circles) on the of Fe-doping concentration of Cr$_2$B. The dotted black and the dashed blue lines are a guides to the eye. A transition is seen from a paramagnetic non-Fermi liquid metal (NFL, yellow shading) with predominant antiferromagnetic correlations to a ferromagnetic metal (FM, blue shading) at the critical concentration  $x_{\rm c}=2.5$\%.}
\end{figure}

The non-Fermi-liquid behavior is revealed through the strong temperature dependence of the $^{11}$B Knight-shift found across the entire phase diagram (Fig. \ref{fig6}). All samples surprisingly show a Curie-Weiss-like dependence of $K_{\rm s}$ at high temperatures. One possible explanation for such a dependence is the presence of localized states in the samples, originating either from defects (i.e. a slight non-stoichiometry in case of the parent Cr$_2$B) or from an orbitally selective Mott transition\cite{Capone, FST} in this multi-band system.  However, a crossover to a low-temperature state where $K_{\rm s}$ follows $T^{-n}$ with a power-law exponent $n\leq 1/2$ contradicts both these possibilities. Rather we conclude that the Fe-doped Cr$_2$B materials family is indeed close to a QCP as originally suggested.\cite{LSchoop} A very similar temperature dependence of the Knight-shift has been found in other archetypal materials close to a QCP, where such dependence was attributed to the existence of strong spin correlations in the metallic state.\cite{CFNMR} 

The second important finding relates to the origin of spin fluctuations and their evolution with Fe-doping. The analysis of the $^{11}$B spin-lattice relaxation data is consistent with ferromagnetic fluctuations. These ferromagnetic fluctuations coexist with antiferromagnetic correlations deduced from the negative Curie-Weiss temperature.  Fe-doping does not significantly affect the ferromagnetic fluctuations. On the other hand, the Fe-dependence of $T_{\rm CW}$ (Fig. \ref{fig6})   suggests that the antiferromagnetic correlations gradually vanish as the doping level approaches the critical Fe concentration. For higher Fe-doping levels, ferromagnetic correlations prevail and as a result the magnetic ordering temperature monotonically and rapidly increases with $x$. 
{\red What remains to be answered in future work is why Fe-doping affects the ferromagnetic correlations to a much lesser degree than the antiferromagnetic correlations.}

In conclusion, we have systematically investigated the effect of Fe-doping on the magnetism in the intermetallic compound Cr$_2$B. $^{11}$B NMR and ESR data suggest that these materials may indeed be close to a quantum critical point at the critical Fe-doping level of $x_{\rm c}\approx 2.5$\%. The data also reveal that antiferromagnetic and ferromagnetic correlations coexist in these materials, but are differently affected by Fe-doping. At $x_{\rm c}$, ferromagnetic correlations prevail and  magnetic ordering is observed at higher doping levels. Our characterization thus indicates that intermetallic Cr$_2$B as an interesting material system where the effect of weak quenched disorder on a ferromagnetic quantum phase transition can be systematically studied.


%
\section*{Acknowledgments}
%
%
D.A. and C.F. acknowledge the financial support by the European Union FP7-NMP-2011-EU-Japan project LEMSUPER under contract no. 283214. D.A. also acknowledges discussions with Peter Jegli\v c about the $^{11}$B NMR spectra. D.A. and C.F. also acknowledge discussions with Peter Adler regarding the magnetic ground state. The research at Princeton University was supported by the US Depaertment of energy, grant DE-FG02-98ER45706.  
\bibliography{Cr2-2015-bib}
\bibliographystyle{apsrev}

\end{document}